\documentclass[12pt]{article}
\usepackage{subfigure}
\usepackage{amssymb,amsmath}
\usepackage{graphicx}
\usepackage{float}
\usepackage{color}
\usepackage[colorlinks=true
,urlcolor=blue
,citecolor=blue
,linkcolor=blue
,pagecolor=blue
,linktocpage=true
,pdfproducer=medialab
]{hyperref}
\usepackage[a4paper]{geometry}
\usepackage{cite}
\usepackage{placeins}
\usepackage[utf8]{inputenc}
\usepackage{cancel}
\usepackage{arydshln}
\makeatletter \renewcommand{\@dotsep}{10000} \makeatother
\usepackage{indentfirst}

\mathchardef\mhyphen="2D

\newcommand{\beq}{\begin{equation}}
\newcommand{\eeq}{\end{equation}}
\newcommand{\bea}{\begin{eqnarray}}
\newcommand{\eea}{\end{eqnarray}}

\def\locald{\rho_0}

\def\mnucl{m_{N}}
\def\redN{\mu_N}
\def\redN{\mu_N}
\def\sigmaN{\sigma}

\newcommand{\mchi}{m_{\chi}}

\newcommand\lsim{\mathrel{\rlap{\lower4pt\hbox{\hskip1pt$\sim$}}
		\raise1pt\hbox{$<$}}}
\newcommand\gsim{\mathrel{\rlap{\lower4pt\hbox{\hskip1pt$\sim$}}
		\raise1pt\hbox{$>$}}}
\def\lsim{\mathrel{\raise.3ex\hbox{$<$\kern-.75em\lower1ex\hbox{$\sim
				$}}}}
\def\gsim{\mathrel{\raise.3ex\hbox{$>$\kern-.75em\lower1ex\hbox{$\sim
				$}}}}

\usepackage[nodisplayskipstretch]{setspace}

\begin{document}

\begin{titlepage}

\begin{flushright}

\end{flushright}
\pagestyle{empty}

\vspace*{0.2in}
\begin{center}
{\Large \bf   Characterisation  of Dark Matter in Direct Detection  \\[0.25cm]
 Experiments: Singlino Versus Higgsino
  }\\
\vspace{1cm}
{\bf  Ya\c{s}ar Hi\c{c}y\i lmaz$^{a,b}$\footnote{E-mail: Y.Hicyilmaz@soton.ac.uk} and Stefano Moretti$^{a}$\footnote{E-mail: S.Moretti@soton.ac.uk} }
\vspace{0.5cm}
\begin{flushleft}
{\it $^a$School of Physics and Astronomy, University of Southampton, Highfield, Southampton SO17 1BJ, United Kingdom} \\
{\it $^b$Department of Physics, Bal\i kesir University, TR10145, Bal\i kesir, Turkey}\\

\end{flushleft}

\end{center}

\vspace{0.5cm}
\begin{abstract}

\end{abstract}
We show how the material used in direct detection experiments of Dark Matter (DM), in the presence of a signal of it, can afford one with the possibility of extracting the nature of the underlying candidate. We do so for the case of a $U(1)'$ Supersymmetric Standard Model (USSM) of E$_6$ origin, by exploiting benchmark points over its parameter space that yield either a Singlino- or Higgsino-like neutralino as DM candidate, the latter being defined in presence of up-to-date constraints, from low to high energy and from collider to non-collider experiments.  However, as our method is general, we  also introduce a model-independent description of our analysis, for the purpose of aiding similar studies in other Beyond the Standard Model (BSM) scenarios. This has been made possible by adapting a rather simple $\chi^2$ analysis normally used for signal extraction in direct detection experiments and the procedure has been applied to Xenon, Germanium and Silicon detectors, those showing maximal and complementary sensitivity to gauge- and Higgs-portal induced interactions of DM with their nuclei.

\end{titlepage}

\section{Introduction}
\label{sec:intro}

Understanding the nature of Dark Matter (DM) is one of the most fundamental problems of particle physics and astronomy. Today, the vast majority of particle physicists and astronomers believe that more than 20 percent of the mass that exists in our universe is composed of non-luminous matter, indeed, called DM. Since 1933, when the astronomer Fritz Zwicky first observed an anomaly in the rotation of a galaxy in the Coma cluster \cite{Zwicky}, which could only be accounted for by such a new form of matter, no direct information is available about DM (e.g., its mass, spin, composition, the symmetry responsible for its stable structure, how it interacts with the Standard Model (SM) particles, etc.). Further, the existence of DM is one of the most obvious reason to seek Beyond SM (BSM) physics, simply because the SM has no DM candidate. Today, observations on the cosmological scale allow us to quantify the abundance of  DM in the universe. Due to the observed abundance, the majority of  DM cannot be in baryonic form, since the Big Bang Nucleosynthesis (BBN) puts an upper bound on the density of baryons in the universe \cite{Schramm:1982nf}. In that case, some  BSM scenarios, like Supersymmetry, can provide non-baryonic DM candidates, so-called Weakly Interacting Massive Particles (WIMPs), which are stable, massive and weakly interacting with the ordinary matter. WIMPs are particularly attractive as they can give  the right amount of relic abundance in the universe measured by Planck \cite{Ade:2015xua} and WMAP \cite{Bennett:2012zja}. Moreover, WIMPs are experimentally appealing DM candidates because of the possibility of their detection. 

Currently, experimental searches for WIMPs as DM are performed under three main approaches. The first of these is searches carried out at colliders, like the Large Hadron Collider (LHC) at CERN. The others are cosmological searches in which the effects of DM are observed directly or indirectly on or under the Earth's surface as well as in space.  Searches for DM at colliders assume  DM production which relies upon the existence of interactions between the  (B)SM  particles and the DM particles. Such DM particles pass invisibly through the detector so that the main search channels are events with missing energy, stemming from a collisions in which a part of the energy goes to undetected particles, which could indeed be WIMPs.

Indirect detection of DM is based on the idea that WIMPs may self-annihilate into SM particles as a flux of cosmic rays, $\gamma$-rays,, neutrinos and/or antimatter which can appear as an excess over the expected astrophysical background. Such an excess is expected to be detected at cosmic rays detectors \cite{Alvisi:1999zk}, $\gamma$-ray telescopes \cite{Ackermann:2015zua} or neutrino observatories \cite{Sullivan:2013zaa}.

Direct detection of DM, on which we focus in this work, aims at detecting WIMPs via the nuclear recoils that arise from an elastic scattering of a WIMP on a target nucleus. Such scatterings occur in the framework of weak interactions, mostly with the exchange of a $Z$ boson, the so-called Spin-Dependent (SD) scattering, and the exchange of a Higgs boson, the so-called Spin-Independent (SI) scattering,  between the WIMP and the nuclei of the target material\footnote{In Supersymmetry (SUSY), sfermion exchanges can contribute to the WIMP-nucleus scattering. However, their contributions are typically small, even for light squarks.}. So,  the rate of the interactions is extremely small and it is needed to have low-background detectors, which are generally placed underground to shield from the cosmic ray background. There are many direct detection experiments  worldwide, in which different type of nuclear targets are used with various background subtraction techniques \cite{Aprile:2018dbl,Cui:2017nnn,Lee:2014zsa,Agnese:2017jvy,Agnese:2017njq,Behnke:2012ys,Behnke:2016lsk,Amole:2015pla,Amole:2017dex,Angloher:2015ewa,Yang:2017yaw,Aguilar-Arevalo:2016ndq,Akerib:2013tjd}.

From a phenomenological perspective, it is important to determine the type of WIMP DM which satisfies the possible signal observed in any direct detection experiment. In order to compare potential WIMP candidates in the same or different theoretical models, we need to know whether different direct detection signals of the DM candidates are seperable. The present work is devoted to the analysis of direct detection signals which belong to two different type of neutralino DM candidates emerging in the same SUSY model, the latter being of E$_6$ origin with a low energy appearance typical of a $U(1)'$ Supersymmetric Standard Model (UMSSM). We use Singlino- and Higgsino-like  neutralino solutions, which satisfy all current experimental bounds, as studied in detail in our previous work \cite{Frank:2020pui}. This present analysis aims at explaining to what extent such two kind of neutralinos are separable for different exposures and target materials in future direct detection experiments. We also observe the effects of the SI and SD cross sections on the discrimination of DM direct detection signals using model-independent benchmarks. 

The organisation of the paper is the following. We will focus on DM direct detection in Section \ref{sec:directdetection}. Then, we present our results in Section \ref{sec:results}. Finally, we summarise and conclude in Section \ref{sec:conclusion}.

\section{Direct Detection of Dark Matter}
\label{sec:directdetection}

As mentioned, the idea of DM direct detection relies on the observation of the nuclear recoil caused by the weak interactions of WIMPs with the nuclei in detector materials\cite{Goodman:1984dc}.  The recoil energy of a nucleus after a collision with a DM particle can be written as:
\begin{equation}
{E_R}=\frac{\redN^2v^2(1-\cos\theta_{\rm CM})}{m_N} \,,
\label{E_r}
\end{equation}
where $ \redN= \frac{\mchi m_N}{\mchi+m_N}$ is the so-called WIMP-nucleus reduced mass, with $ \mchi  $ and $ m_N $ the masses of the WIMP and nucleus, respectively. Here, $ v $ is the relative velocity between the WIMP and nucleus while $ \theta_{\rm CM}  $ is the scattering angle in the Center-of-Mass (CM) frame. Eq. (\ref{E_r}) shows that the amount of transferred energy $ E_R $  to the nucleus by the WIMP depends on the scattering angle in the CM frame as well as the two masses and relative velocity. This energy is maximal in the case the WIMP backscatters, i.e., $ \theta_{\rm CM}=\pi $, while there is no transferred energy if the WIMP passes through the detector without interaction, i.e., $ \theta_{\rm CM}=0 $.  So, the maximal recoil energy of a nucleus scattered by a WIMP for a given velocity $ v $ is obtained as:
\begin{equation}
{E_{\rm max}(v)}=\frac{2\redN^2v^2}{m_N}\,.
\label{E_max}
\end{equation}
Moreover, for an elastic scattering,  the minimal velocity $ v_{\rm min} $ for a WIMP mass $ \mchi $ to be able to induce a nuclear recoil of energy $ E_R $ is
\begin{equation}
{v_{\rm min}(E_R)}=\sqrt{\frac{E_R m_N}{2\redN^2}}  \,.
\label{v_min}
\end{equation}

In Eq. (\ref{E_r}), if we express the recoil energy $ E_{R} $ as a function of the mass of the target nucleus $ m_N $ for any given DM mass $ \mchi $, with velocity $ v $ and $ \theta_{\rm CM}=\pi $, it can be seen that $ E_{R} $ is maximal in the case of $ m_N=\mchi $, which satisfies the equation $ dE_R/dm_N=0 $. This means that the maximal transfer of energy, i.e., the maximal recoil energy, for a given WIMP mass takes place when using a target nucleus with mass approximately equal to that of the WIMP. As a consequence, the sensitivity of a detector increases as the mass of the nucleus to be used as a target in the detector get closer to the mass of the WIMP to be probed in the detector. Heavier nuclei will give a detector more sensitive to heavier WIMPs. In fact, other than for discovery, the detector sensitivity is also important for the differentiation of possible signals, so a dedicated detector (i.e., with suitable material chosen) could be designed in responsive mode to a discovery in another medium. 

The differential event rate, called  the nuclear recoil spectrum or the nuclear recoil energy distribution, expressed in terms of the number of events per unit energy per unit time per unit target material mass (in general /keV/kg/day, referred to as a differential rate unit) for a WIMP with mass $\mchi$ and a nucleus with mass $\mnucl$
is given by \cite{Lewin:1995rx}
\begin{equation}
\frac{dN}{dE_R}=\frac{\locald}{\mnucl\,\mchi}\int_{v>v_{\rm min}} v
f(v) \frac{d\sigmaN}{dE_R}(v,E_R)\, d v\,,
\label{diff_rate}
\end{equation}
where $\rho_0$ is the local WIMP density,
$\frac{d\sigmaN}{dE_R}(v,E_R)$ is the differential cross section for
the WIMP-nucleus elastic scattering and $f(v)$ is the WIMP speed
distribution in the detector frame, for which we assume a standard isotropic Maxwellian at rest. The lower limit of the integration over WIMP speed $ v_{\rm min} $ is equal to the minimal WIMP velocity shown in Eq. (\ref{v_min}) while the upper limit is informally the local escape speed, $ v_{\rm esc}  $,  the maximum speed in the Galactic rest frame for WIMPs which are gravitationally bound to the Milky Way. The total number of recoil events (per kilogram per day) can be found by integrating the differential event rate over all the possible recoil energies: 
\begin{equation}
N=\int_{E_T}^{E{\rm max}}
dE_R\frac{\locald}{\mnucl\,\mchi}\int_{v>v_{\rm min}} 
v f(v) \frac{d\sigmaN}{dE_R}(v,E_R)\, d v \,,
\label{tot_rate}
\end{equation}
where $E_{T}$ is the threshold energy, the smallest recoil energy
which the detector is capable of measuring, and $E_{\rm max}$ is the maximal recoil energy expressed in Eq. (\ref{E_max}).

The differential scattering cross section $\frac{d\sigmaN}{dE_R}(v,E_R)$ includes different types of WIMP-nucleus interactions. These interactions mainly depend on the WIMP-quark interaction strength, however, the resulting cross section is translated into the  WIMP-nucleon cross section by using hadronic matrix elements which describe the nucleon content in terms of quarks and gluons. As intimated, two types of interaction are considered \cite{Goodman:1984dc}: the spin-spin interaction (SD), where the WIMP couples to the spin of the nucleus by the exchange of a $Z$ boson, and the scalar interaction (SI), where the WIMP couples to the mass of the nucleus by the exchange of a Higgs boson. In this work, we analyse the combination of both interactions:
\begin{equation}
\frac{d\sigmaN}{dE_R}= \left(\frac{d\sigmaN}{dE_R}\right)_{\rm SI}
+\left(\frac{d\sigmaN}{dE_R}\right)_{\rm SD}\,.
\label{sigma_tot}
\end{equation}

The contributions of the SI WIMP-nucleus scattering result from the scalar and vector interaction terms in the Lagrangian. The presence of these, terms shown in Eq. (\ref{lag_SI}), which include the couplings between SM particles and WIMP candidate, directly depends upon the  particle physics model (see Ref. \cite{Jungman:1995df} for Feynman diagrams), 

\begin{equation}
{\cal L}_{\rm SI} = \alpha_q^S
\bar\chi\chi \bar qq +
\alpha_q^V\bar\chi\gamma_\mu\chi\bar q\gamma^\mu q
\,.
\label{lag_SI}
\end{equation}
The SI differential cross section can be written as:
\begin{equation}
\left(\frac{d\sigmaN}{dE_R}\right)_{\rm SI}=\frac{\mnucl \sigma_0^{\rm SI}
	F^2(E_R) }{2\redN^2v^2} \,,
\label{XS_SI}
\end{equation}
where $F^2(E_R)$ is the nuclear form factor for SI interactions  which is a Fourier transform of the nucleon density and parametrised in terms of the momentum transfer as \cite{Engel:1991wq}:
\begin{equation}
F^2(q)=\left(\frac{3j_1(qR_1)}{qR_1}\right)^2\,
\exp\left[-q^2s^2\right] \,,
\label{SI_form}
\end{equation}
where $j_{1}$ is a spherical Bessel function, $s\simeq 1$~fm is a measure of the nuclear skin thickness and $R_1=\sqrt{R^2-5s^2}$ with $R\simeq 1.2\, A^{1/2} \, {\rm fm}$, $A$ being the mass number of the nucleon.  The form factor is normalised to
unity at zero momentum transfer, $F(0)=1$. Here, $ \sigma_0^{\rm SI} $ is the SI zero momentum WIMP-nucleus cross section and leads to the following expression: 
\begin{equation}
\sigma_0^{\rm SI}=\frac{4 \redN^2}{\pi} \left[Z f^p + (A-Z) f^n\right]^2 \,,
\label{sigma_zero_SI}
\end{equation}
where $ Z $ is the nucleus atomic number, $ f_p $ and $ f_p $ are the WIMP-proton  and  WIMP-neutron couplings, respectively. In most cases the WIMP coupling to neutrons and protons is very similar, $f^p\approx f^n$, and  Eq. (\ref{sigma_zero_SI})  can be expressed as 
\begin{equation}
\sigma_0^{\rm SI}=\frac{4 \redN^2 A^2 f^p}{\pi}  \,.
\label{sigma_zero_SI_}
\end{equation}

The SD scattering is due to the interaction of a WIMP with the spin of the nucleus through the part of the Lagrangian given by the axial-vector interaction terms such as
\begin{equation}
{\cal L}_{\rm SD} =  \alpha_q^A (\bar\chi\gamma^\mu\gamma_5\chi)
(\bar q\gamma_\mu\gamma_5 q) \,.
\end{equation}
The SD differential cross section is \cite{Pato:2011de}
\begin{equation}
\left(\frac{d\sigmaN}{dE_R}\right)_{\rm SD}=\frac{16 \mnucl}{\pi
	v^2}\Lambda^2 G_F^2 J(J+1) \frac{S(E_R)}{S(0)} \,,
\label{XS_SD}
\end{equation}
where $ F^2_{\rm SD}(E_R)=\frac{S(E_R)}{S(0)} $ is the SD form factor, which depends on the spin structure of a nucleus. Furthermore, $ \Lambda = (a^n \left\langle S_n\right\rangle + a^p \left\langle S_p\right\rangle )/J $, where $ J $ is the spin of the target nucleus, $ a^p $($ a^n $) is the axial WIMP-proton(-neutron) coupling and $ \left\langle S_p\right\rangle $($ \left\langle S_n\right\rangle $) is the expectation value of the spin of protons(neutrons) in the nucleus. The SD zero momentum cross section $ \sigma_0^{\rm SD} $ can be expressed as:
\begin{equation}
\sigma_0^{\rm SD} =\frac{32}{\pi}\redN^2\Lambda^2 G_F^2 J(J+1) \,.
\label{sigma_zero_SD_}
\end{equation}
As seen from Eqs. (\ref{sigma_zero_SI_}) and (\ref{sigma_zero_SD_}),  the SI contribution is  directly proportional to the square of the number of nucleons, $A^2$, whereas the SD one is a function of the nuclear angular momentum, $(J+1)/J$.

Finally, the total WIMP-nucleus cross section can be calculated by adding the gauge and scalar components shown in Eqs. (\ref{XS_SI}) and (\ref{XS_SD}). The form factor, $F(E_R)$, encodes the
dependence on the momentum transfer, $q=\sqrt{2 \mnucl E_R}$, 
and accounts for the coherence loss which leads to a suppression in the event rate
for heavy WIMPs or nucleons. We can rewrite the differential cross section in Eq. (\ref{sigma_tot}) as
\begin{equation}
\frac{d\sigmaN}{dE_R}=\frac{\mnucl}{2\redN^2v^2}\left(\sigma_0^{\rm SI}
F^2_{\rm SI}(E_R) + \sigma_0^{\rm SD} F^2_{\rm SD}(E_R)\right).
\label{sigma_tot2}
\end{equation}

In this work, we employ MicrOMEGAs (version 5.0.8) \cite{Belanger:2018ccd} to calculate the nuclear recoil spectrum shown in Eqs. (\ref{diff_rate}) and (\ref{tot_rate}). For the discrimination between the signal and  background, it is needed to calculate the variance $\chi^2$ \cite{Bernal:2009tt,Profumo:2009tb}:
\begin{equation}
\chi^2=
\sum_{i=1}^n\left(\frac{N_i^{\rm tot}-N_i^{\rm bkg}}{\sigma_i}\right)^2\ ,
\end{equation}
where
$N^{\rm tot}=N^{\rm sign}+N^{\rm bkg}$ is the total rate  measured by the detector,
with $N^{\rm sign}$ and $N^{\rm bkg}$  the signal and  background yields, respectively.  However, we can also use this same $\chi^2$  analysis to separate  different type of signals in the case that we assume, e.g.,  one type of them being the nuclear recoil distribution for Higgsino-like neutralino (as signal) and the other one being the nuclear recoil distribution for Singlino-like neutralino (as background). 

Of course, this presumes a signal being detected. With this in mind, we will divide the energy  range between $5$ and $50$ keV in $n=9$ equidistant energy bins.   Here, we assume a Gaussian error $\sigma_i=\sqrt{\frac{N_i^{tot}}{M\cdot T}}$ on the measurement, where $M$ is the detector mass and $T$ the exposure time. We require $ \chi^2 > 15.51 $ to separate two direct detection signals at  the 95\% Confidence Level (CL). In our work we calculate $\chi^2$ values and probe the discrimination of the direct detection signals, which results from different type of DM candidates, for four different exposures, 2 $ t\cdot y $, 6  $ t\cdot y $, 20  $ t\cdot y $ and 200  $ t\cdot y $.  Especially, the exposures of  20  $ t\cdot y $ and 200  $ t\cdot y $ are the maximum expected  exposures for the next generation direct detection experiments,  XENONnT/LUX-ZEPLIN \cite{Aprile:2015uzo,Mount:2017qzi} and DARWIN \cite{Aalbers:2016jon}, respectively.

\section{Results}
\label{sec:results}

In this section, we will  present the results of our analysis on the comparison of DM direct detection signals, in presence of 
various nuclei, 
 for different kinds of DM candidates in model-dependent and model-independent frameworks. In order to do so, we have selected some benchmarks in both frameworks, shown in Tables \ref{dptBench} and \ref{indBench}, respectively. The model-dependent benchmarks are selected from the UMSSM model results worked out in Ref. \cite{Frank:2020pui} while the model-independent ones, which do not result from any physical model, are created to compare the direct detection signals by varying arbitrarily the values of the SI and SD cross sections. It can be noted that we tried to choose  model-independent benchmarks which have  cross section and WIMP mass values similar to those of the model-dependent ones. 

\begin{table}[h]
	\centering
	\fontsize{11}{11}\selectfont
	\begin{tabular}{c||cccc}
		\hline
		\hline
		DM benchmark & $m_{\chi_{1}^{0}}\,[\textrm{GeV}]$ & $\sigma_n^{\rm SI}\,[\textrm{pb}]$ & $\sigma_n^{\rm SD}\,[\textrm{pb}]$ & Composition$\,\textrm{[\%]}$  \\
		\hline
		\textbf{BM-DPT I} & 1134 & 1.07x$10^{-11}$ & 1.06x$10^{-6}$ & Higgsino-like, $ 98\% $  \\
		\textbf{BM-DPT II} & 1181 & 1.02x$10^{-11}$ & 1.67x$10^{-6}$ & Singlino-like, $ 96\% $  \\
		\textbf{BM-DPT III} & 1161 & 7.92x$10^{-12}$ & 1.83x$10^{-6}$ & Singlino-like, $ 96\% $  \\
		\textbf{BM-DPT IV} & 1013 & 7.83x$10^{-12}$ & 2.21x$10^{-6}$ & Singlino-like, $ 97\% $  \\
		\textbf{BM-DPT V} & 1135 & 2.01x$10^{-11}$ & 1.49x$10^{-6}$ & Singlino-like, $ 95\% $  \\
		\textbf{BM-DPT VI} & 1114 & 3.41x$10^{-11}$ & 1.54x$10^{-6}$ & Singlino-like, $ 95\% $  \\		
		\hline
	\end{tabular}
	\caption{The model-dependent DM benchmarks selected from the results shown in Figure 7 of Ref.  \cite{Frank:2020pui} and used to generate direct detection signals.}
	\label{dptBench}
\end{table}

As seen from Table \ref{dptBench}, there is only one  Higgsino-like DM benchmark \textbf{BM-DPT I}, whereas there are five Singlino-like DM benchmarks with various SI and SD cross sections. We will assess whether the described $\chi^2$ analysis can enable us to separate the former from the latter. Therefore,  \textbf{BM-DPT I} can be regarded as our signal (in the sense discussed above) while the others are backgrounds. They all corresponds to actual {\sl discrete} parameter space points in the UMSSM.  Conversely,  the model-independent benchmarks shown in Table \ref{indBench} are used to describe the typical {\sl continuous} behaviour over the relevant recoil spectra.

\begin{table}[H]
	\centering
	\fontsize{11}{11}\selectfont
	\begin{tabular}{c||ccc}
		\hline
		\hline
		DM benchmark & $m_\chi\,[\textrm{GeV}]$ & $\sigma_n^{\rm SI}\,[\textrm{pb}]$ & $\sigma_n^{\rm SD}\,[\textrm{pb}]$   \\
		\hline
		\textbf{BM-IND I} & 1000 & 1.0x$10^{-11}$ & 1.0x$10^{-6}$  \\
		\textbf{BM-IND II} & 1000 & 1.0x$10^{-11}$ & 1.5x$10^{-6}$  \\
		\textbf{BM-IND III} & 1000 & 1.0x$10^{-11}$ & 2.0x$10^{-6}$  \\
		\textbf{BM-IND IV} & 1000 & 1.0x$10^{-11}$ & 2.5x$10^{-6}$  \\
		\textbf{BM-IND V} & 1000 & 1.5x$10^{-11}$ & 1.0x$10^{-6}$  \\
		\textbf{BM-IND VI} & 1000 & 2.0x$10^{-11}$ & 1.0x$10^{-6}$  \\ 		
		\textbf{BM-IND VII} & 1000 & 2.5x$10^{-11}$ & 1.0x$10^{-6}$  \\
		\hline
	\end{tabular}
	\caption{The model-independent DM benchmarks used to generate direct detection signals.}
	\label{indBench}
\end{table}

In this analysis, the direct detection signals of the benchmarks shown in Tables \ref{dptBench} and \ref{indBench} are calculated as the differential event rate in /keV/kg/day   for a Xenon, Germanium and Silicon target nucleus. Each detector composed of these targets surely has different experimental backgrounds. The experimentalists deal with reducing these backgrounds in  direct detection experiments to increase sensitivity. In this work, however, we assume that our signals are larger than the experimental backgrounds since we intend to focus on the discrimination between the expected direct detection signals of two different DM candidates, one Higgsino- and the other Singlino-like. In the forthcoming figures, for each type of target, the model-independent differential event distributions are displayed in large plots in the top panels. In these plots, a black line refers to the distribution for the benchmark \textbf{BM-IND I} while the rates for other benchmarks are tagged as coloured lines. The small plots in the top panels indicate instead the normalisation of the coloured lines to the black signal with the same colour coding. In the model-dependent case shown in the bottom panels, the colour coding is same as with the model-independent case and we show only the plots with the normalisation of coloured rates to black signal (again, corresponding to the benchmark \textbf{BM-DPT I}). In both casess, the right (left) panels display the different recoil shapes of signals according to varying SI (SD) cross sections. 

\begin{figure}[!t]
	\subfigure{\includegraphics[scale=0.4]{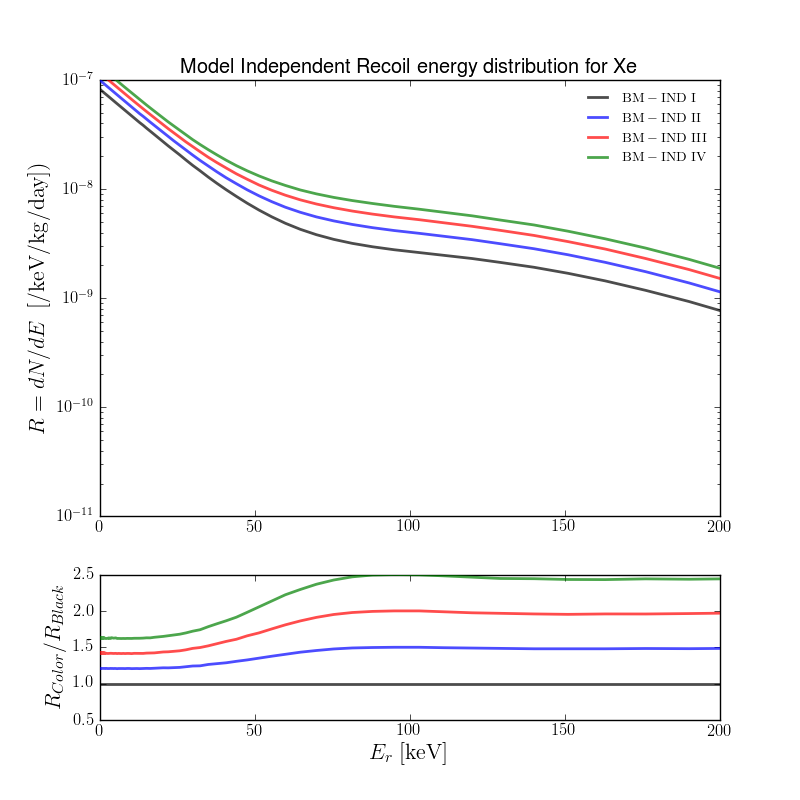}}
	\subfigure{\includegraphics[scale=0.4]{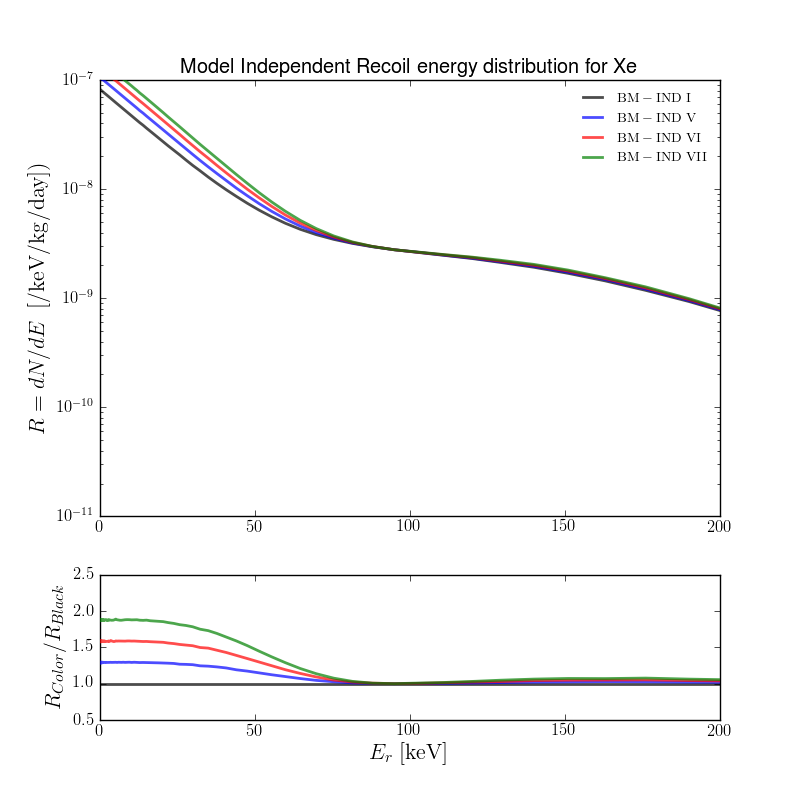}}
	\subfigure{\includegraphics[scale=0.4]{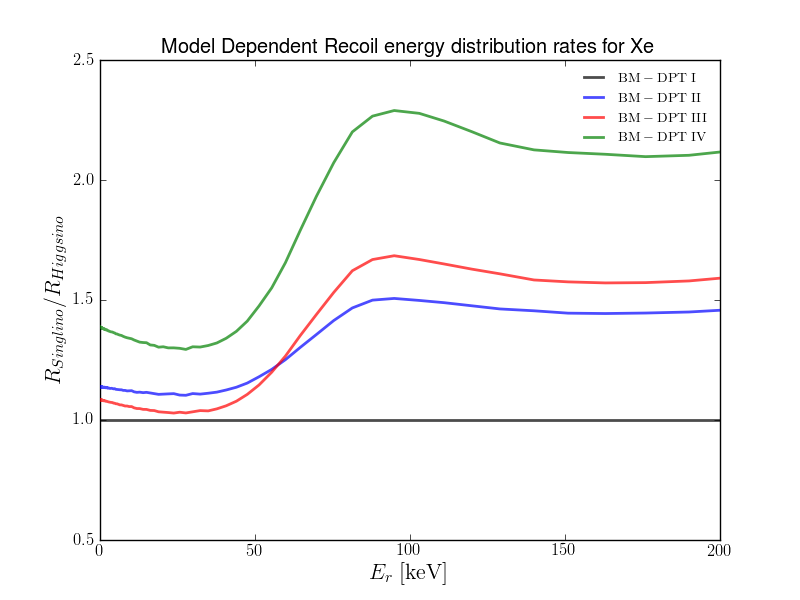}}
	\subfigure{\includegraphics[scale=0.4]{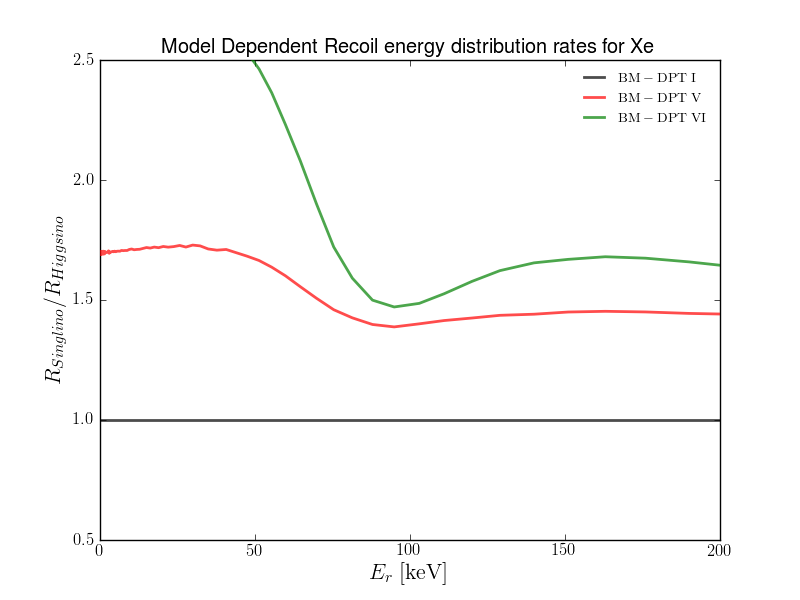}}
	\caption{The direct detection signals of the benchmarks shown in Tables \ref{dptBench} and \ref{indBench} as a function of the nucleus recoil energy ($E_r\equiv E_R$) for a Xenon target. The model-independent differential event distributions are displayed in the large plots of the top panels. In these plots, a black line refers to the benchmark \textbf{BM-IND I} while the other benchmarks are tagged in coloured lines. The small plots of the top panels indicate the normalisation of the coloured benchmarks relative to the  black one. In the model-dependent case shown in the bottom panels, the colour coding is the same as in the model-independent case but we only show the plots with the relative normalisation of the benchmarks. For both cases, the right(left) panels display the different recoil spectra according to varying SI(SD) cross sections.}
	\label{fig:Er_Event_Xe}
\end{figure}

Figure \ref{fig:Er_Event_Xe} shows the direct detection signals of the benchmarks shown in Tables \ref{dptBench} and  \ref{indBench} for a Xenon target nucleus with the mass number of 131. According to the top left panels, the effect of changing the SD cross section on the difference between black and any coloured signal is small up to 50 keV of the nuclear recoil energy, compared to the rate of changing the SD cross section. For larger recoil energies, the coloured to black signal ratio, shown in the small plots in the top panels,  approximates the ratio between the SD cross sections of two signals. The reason is due to the distribution of the SI nuclear form factor for the Xenon atom shown in Eq. (\ref{SI_form}). According to Ref. \cite{Lewin:1995rx}, the value of the SI nuclear form factor drops significantly for a Xenon target as the nuclear recoil energy increases. Therefore, the contribution of the SI interaction cross section to the event rate shown in Eq. (\ref{sigma_tot2}) decreases rapidly for larger recoil energies, although Xenon has large a mass number of nucleon, $A$, upon which the SI zero-momentum cross section shown in Eq. (\ref{sigma_zero_SI_}) depends. In the case of varying SI cross sections with same DM masses and SD cross sections (top right panel), it can be noted that the aforementioned result is verified, since it seems that there is no difference between the signals with same SD cross sections for larger recoil energy. When we look at the ratio between the model-dependent distributions (bottom panels), it can easily be seen that the variation in the SI cross sections shown in Table \ref{dptBench} largely affects the difference of the direct detection signals for smaller recoil energies than about 50 keV, as the SD cross section becomes important for larger recoil energies.

\begin{table}[H]
	\setstretch{1.5}
	\centering
	\begin{tabular}{|c|c|c||c|c|}
		\hline
			&\multicolumn{2}{|c||}{$\chi^2$ values for top left panel} & \multicolumn{2}{|c|}{$\chi^2$ values for top right panel} \\
		\hline
		$ N_{\rm sign}/N_{\rm bkg} $ & $ \chi^2 (20\, t \cdot y) $ & $ \chi^2 (200\, t \cdot y)$& $ \chi^2 (20\, t \cdot y) $ & $ \chi^2 (200\, t \cdot y)$ \\
		\hline
		Blue/Black & $ 0.97 $ & $ 9.75 $ & $ 1.74 $ & $ 17.49$  \\
		\hline
		Red/Black & $ 3.32 $ & $ 33.25 $& $ 5.71 $ & $ 57.13 $ \\
		\hline
		Green/Black & $ 6.51 $ & $ 65.18 $& $ 10.86 $ & $ 108.62 $ \\
		\hline
		\hline
			&\multicolumn{2}{|c||}{$\chi^2$ values for bottom left panel} & \multicolumn{2}{|c|}{$\chi^2$ values for bottom right panel} \\
		\hline
		$ N_{\rm sign}/N_{\rm bkg} $ & $ \chi^2 (20\, t \cdot y) $ & $ \chi^2 (200\, t \cdot y)$& $ \chi^2 (20\, t \cdot y) $ & $ \chi^2 (200\, t \cdot y)$ \\
		\hline
		Blue/Black & $ 0.39 $ & $ 3.93 $ & $-$ & $-$  \\
		\hline
		Red/Black & $  0.15 $ & $ 1.53 $& $ 6.65 $ & $ 66.55 $ \\
		\hline
		Green/Black & $ 2.51 $ & $ 25.11 $& $ 22.53 $ & $ 225.39 $ \\
		\hline		
	\end{tabular}

	\caption{ $\chi^2$ values for the panels in Figure \ref{fig:Er_Event_Xe}. }
	\label{table_xe_1}
\end{table}

The main energy range for the signal in the direct detection experiments with Xenon target has generally been between the recoil energies of 5 and 50 keV \cite{Aprile:2015uzo}. Looking at the results of Figure \ref{fig:Er_Event_Xe}, it can be concluded that the SI cross sections of the DM candidates play a more significant role than the SD cross sections in determining the type of DM in  direct detection experiments with Xenon target. In Table \ref{table_xe_1}, we display the $\chi^2$ values of the benchmarks shown in Figure \ref{fig:Er_Event_Xe} for the exposures of 20 $ t\cdot y $ and 200 $ t\cdot y $, the proposed maximum exposures for the upcoming direct detection experiments XENONnT/LUX-ZEPLIN \cite{Aprile:2015uzo,Mount:2017qzi} and DARWIN \cite{Aalbers:2016jon}, respectively. According to the table, separating the black benchmark  from the coloured ones with varying SI or SD cross sections and same masses cannot generally be possible for exposure of 20 $ t\cdot y $ (i.e., in the near future) while exposure of 200 $ t\cdot y $ (i.e., in the far future) can easily provide the separation conditions for the various DM assumptions. Certainly, though, the larger gaps between the signals which belong to different DM candidates can be exploited in presence of exposures of 20 $ t\cdot y $, as documented in the last $\chi^2$ result of the  bottom right panel.

\begin{figure}[!t]
	\hspace*{-0.25cm}
	\subfigure{\includegraphics[scale=0.4]{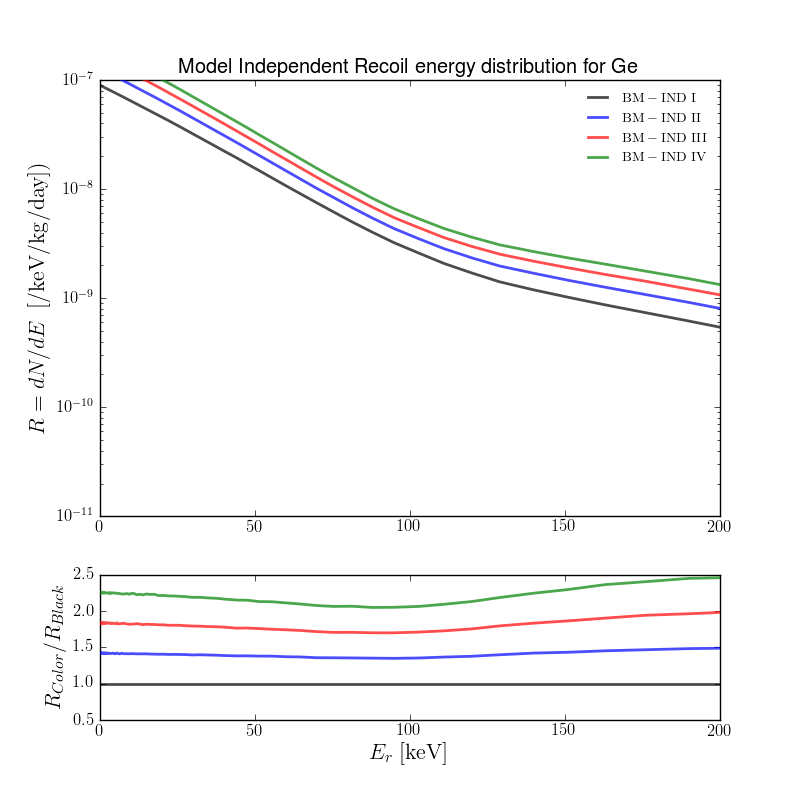}}
	\subfigure{\includegraphics[scale=0.4]{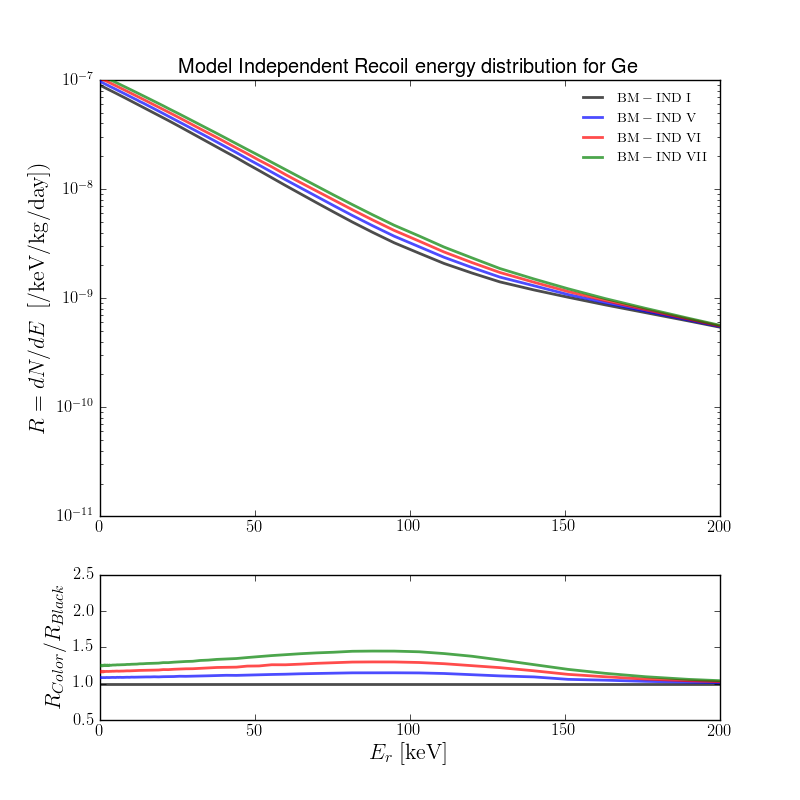}}
	\subfigure{\includegraphics[scale=0.4]{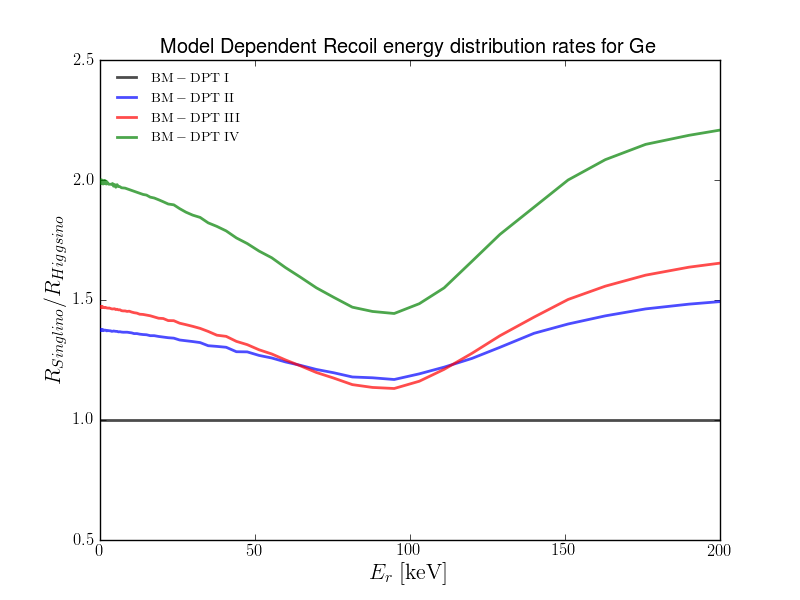}}
	\subfigure{\includegraphics[scale=0.4]{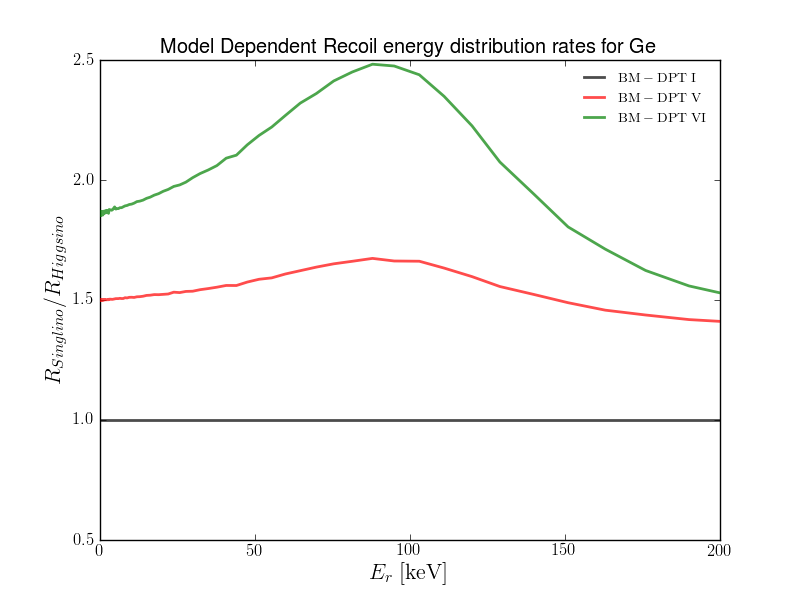}}
	\caption{Same as in Figure \ref{fig:Er_Event_Xe} for Germanium.}
	\label{fig:Er_Event_Ge}
\end{figure}

In Figure \ref{fig:Er_Event_Ge}, we extend the previous analysis to the case of a Germanium target nucleus with a mass number of 73. Unlike the case of a Xenon target nucleus (with a mass number of 131), the effect of the SI cross section on the differentiation of the signals is quite small for the whole range of nuclear recoil energies (top right panel), since the SI zero-momentum cross section shown in Eq. (\ref{sigma_zero_SI_}) is proportional to the square of the atomic mass number (as already mentioned). Beside this, the scaling factor of Germanium for the SD cross sections, $(J+1)/J$, is larger than for Xenon \cite{Schnee:2011ooa}. This also ensures  a event rate at the same level as with Xenon. In short, the difference between the SD cross sections of DM candidates gives a dominant contribution to be able to separate those signals from each other in the direct detection experiments with a Germanium target (top left and bottom left panels). The $\chi^2$ values for the distributions in Figure \ref{fig:Er_Event_Ge}  are presented in Table \ref{table_ge_1} and the most interesting result is that the recoil signals of two DM candidates with same SD cross sections cannot be discriminated in direct detection experiments with a Germanium target even for an exposure of 200 $ t\cdot y $ (these are the $\chi^2$ values for the top right panel).

\begin{table}[H]
	\setstretch{1.5}
	\centering
	\begin{tabular}{|c|c|c||c|c|}
		\hline
		&\multicolumn{2}{|c||}{$\chi^2$ values for top left panel} & \multicolumn{2}{|c|}{$\chi^2$ values for top right panel} \\
		\hline
		$ N_{\rm sign}/N_{\rm bkg} $ & $ \chi^2 (20\, t \cdot y) $ & $ \chi^2 (200\, t \cdot y)$& $ \chi^2 (20\, t \cdot y) $ & $ \chi^2 (200\, t \cdot y)$ \\
		\hline
		Blue/Black & $ 3.59 $ & $ 35.98 $ & $ 0.18 $ & $ 1.85 $  \\
		\hline
		Red/Black & $ 11.12 $ & $ 111.20 $& $ 0.68 $ & $ 6.87 $ \\
		\hline
		Green/Black & $ 20.38 $ & $ 203.84 $& $ 1.44 $ & $ 14.44 $ \\
		\hline
		\hline
		&\multicolumn{2}{|c||}{$\chi^2$ values for bottom left panel} & \multicolumn{2}{|c|}{$\chi^2$ values for bottom right panel} \\
		\hline
		$ N_{\rm sign}/N_{\rm bkg} $ & $ \chi^2 (20\, t \cdot y) $ & $ \chi^2 (200\, t \cdot y)$& $ \chi^2 (20\, t \cdot y) $ & $ \chi^2 (200\, t \cdot y)$ \\
		\hline
		Blue/Black & $ 2.12 $ & $ 21.27 $ & $-$ & $-$  \\
		\hline
		Red/Black & $ 3.13 $ & $ 31.36 $& $ 3.45 $ & $ 34.53 $ \\
		\hline
		Green/Black & $ 10.27 $ & $ 102.76 $& $ 8.23 $ & $ 82.31 $ \\
		\hline		
	\end{tabular}
	
	\caption{ $\chi^2$ values for the panels in Figure \ref{fig:Er_Event_Ge}. }
	\label{table_ge_1}
\end{table}
 
\begin{figure}[!t]
	\hspace*{-0.25cm}
	\subfigure{\includegraphics[scale=0.4]{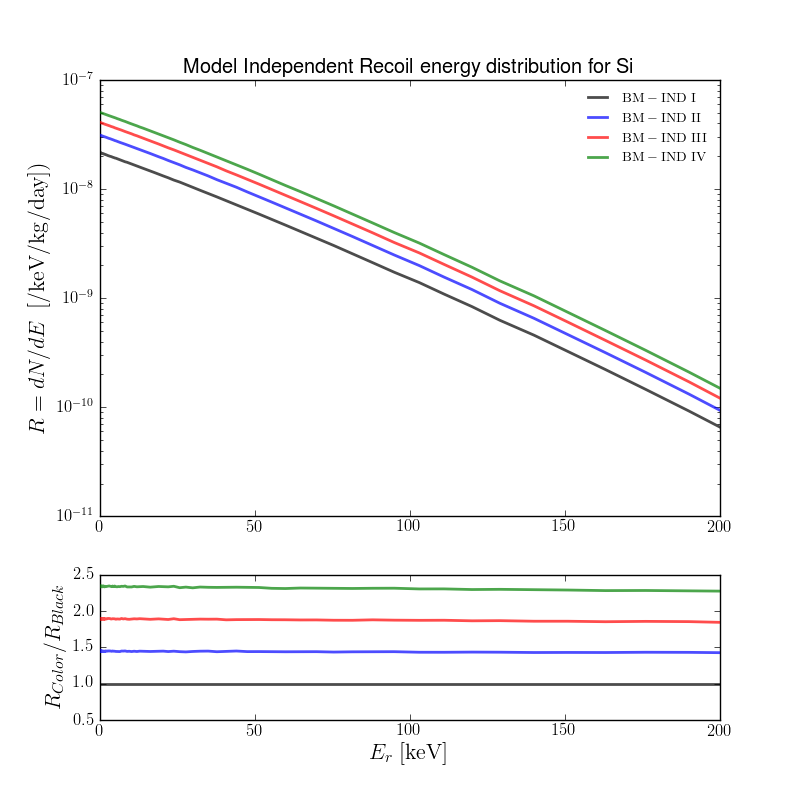}}
	\subfigure{\includegraphics[scale=0.4]{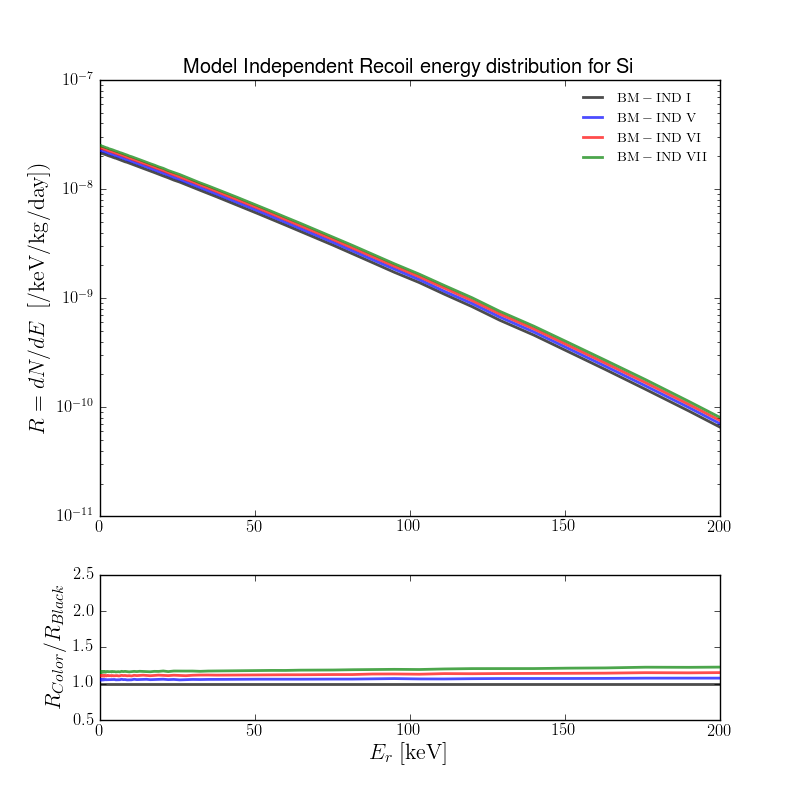}}
	\subfigure{\includegraphics[scale=0.4]{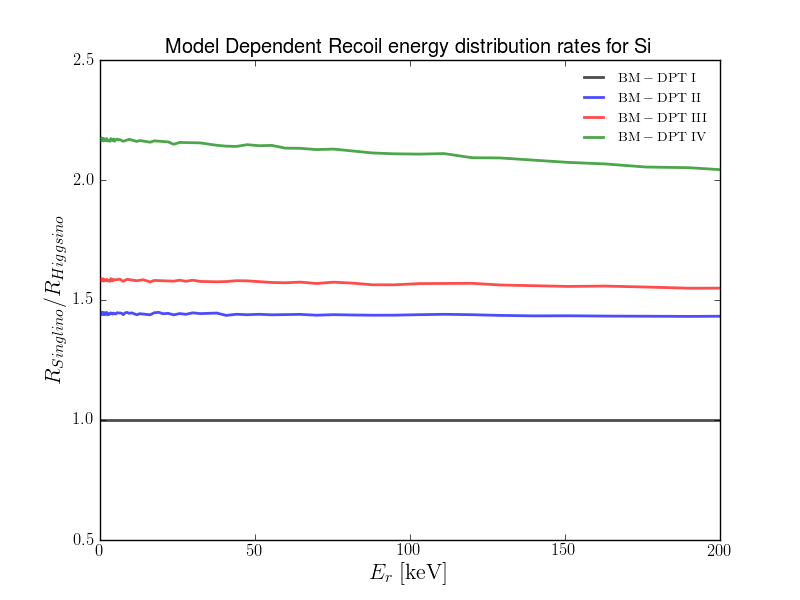}}
	\subfigure{\includegraphics[scale=0.4]{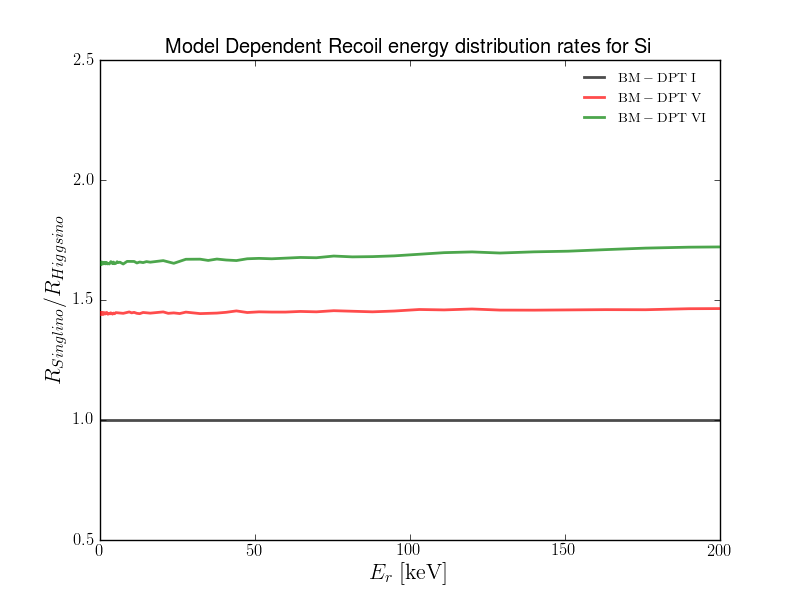}}
	\caption{Same as in Figure \ref{fig:Er_Event_Xe} for Silicon.}
	\label{fig:Er_Event_Si}
\end{figure}

Figure \ref{fig:Er_Event_Si},  showing the nuclear recoil distributions for a Silicon target nucleus with mass number of 29, leads to conclusions similar to those of the Germanium case. However, as shown in Table \ref{table_si_1}, differentiating the recoil distributions of different DM candidates in a Silicon detector is harder, since the event rate is lower than for Germanium due to the smaller mass number and also scaling factor for the SD cross section.

\begin{table}[H]
	\setstretch{1.5}
	\centering
	\begin{tabular}{|c|c|c||c|c|}
		\hline
		&\multicolumn{2}{|c||}{$\chi^2$ values for top left panel} & \multicolumn{2}{|c|}{$\chi^2$ values for top right panel} \\
		\hline
		$ N_{\rm sign}/N_{\rm bkg} $ & $ \chi^2 (20\, t \cdot y) $ & $ \chi^2 (200\, t \cdot y)$& $ \chi^2 (20\, t \cdot y) $ & $ \chi^2 (200\, t \cdot y)$ \\
		\hline
		Blue/Black & $ 0.97 $ & $ 9.72 $ & $ 0.019 $ & $ 0.19 $  \\
		\hline
		Red/Black &  $ 2.97 $ & $ 29.72$& $ 0.074 $ & $ 0.74 $ \\
		\hline
		Green/Black & $ 5.41 $ & $ 54.12 $& $ 0.16 $ & $ 1.60 $ \\
		\hline
		\hline
		&\multicolumn{2}{|c||}{$\chi^2$ values for bottom left panel} & \multicolumn{2}{|c|}{$\chi^2$ values for bottom right panel} \\
		\hline
		$ N_{\rm sign}/N_{\rm bkg} $ & $ \chi^2 (20\, t \cdot y) $ & $ \chi^2 (200\, t \cdot y)$& $ \chi^2 (20\, t \cdot y) $ & $ \chi^2 (200\, t \cdot y)$ \\
		\hline
		Blue/Black & $0.95 $ & $ 9.56 $ & $-$ & $-$  \\
		\hline
		Red/Black & $ 1.50 $ & $ 15.09 $& $ 0.95 $ & $ 9.58 $ \\
		\hline
		Green/Black & $ 4.41 $ & $ 44.18 $& $ 1.80 $ & $ 18.08 $ \\
		\hline		
	\end{tabular}
	
	\caption{ $\chi^2$ values for the panels in Figure \ref{fig:Er_Event_Si}. }
	\label{table_si_1}
\end{table}

\begin{figure}[!t]
	\centering
	
	\begin{minipage}[b]{.5\textwidth}
		\centering
		\includegraphics[scale=0.4]{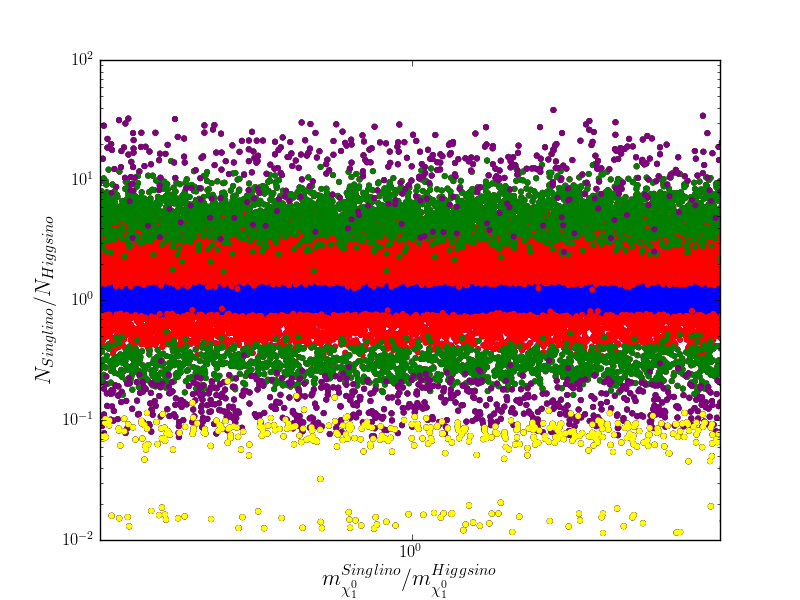}
	\end{minipage}%
	\begin{minipage}[b]{.5\textwidth}
		\centering
		\includegraphics[scale=0.4]{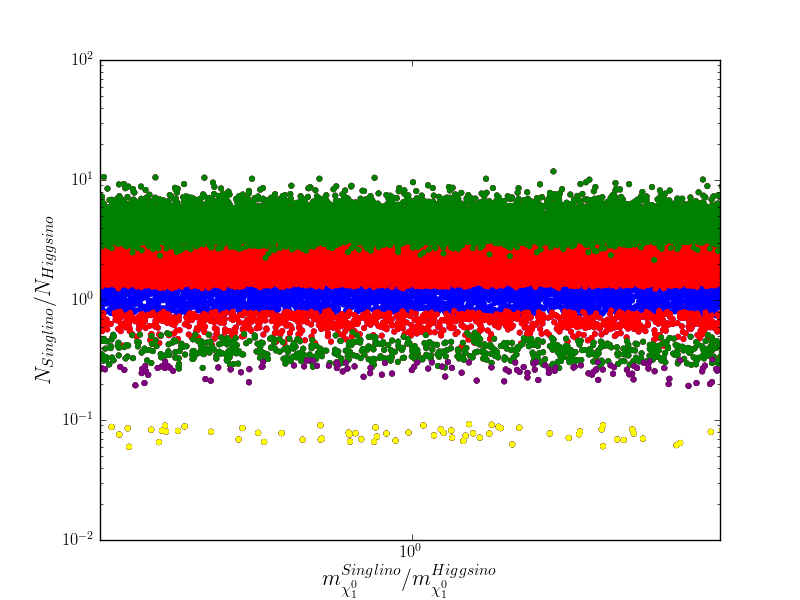}
	\end{minipage}
	
	\bigskip
	
	\begin{minipage}{.5\textwidth}
		\centering
		\includegraphics[scale=0.4]{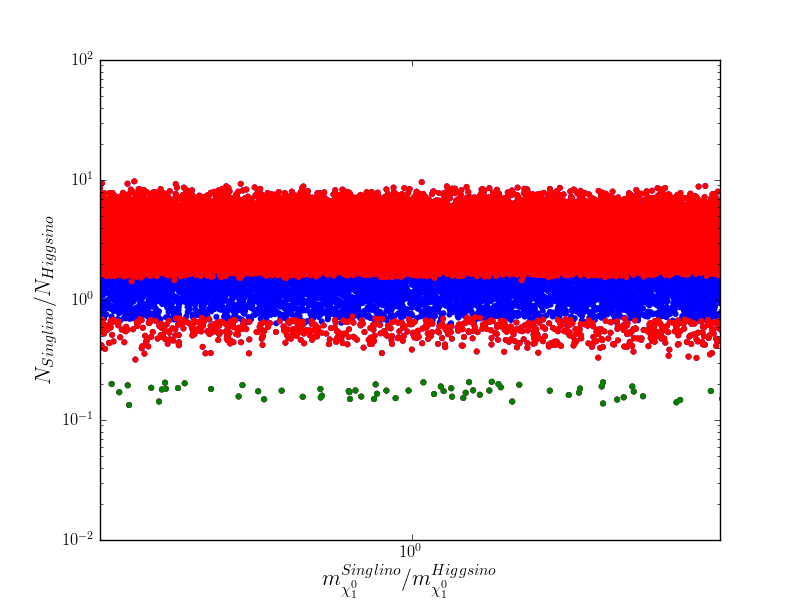}
	\end{minipage}

	\caption{The ratio between Singlino- and Higgsino-like DM masses in the range of [0.99,1.01] versus the ratio between Singlino- and Higgsino-like DM total number of events between 5 and 50 keV nuclear recoil energy for Xenon (top left panel), Germanium (top right panel) and Silicon (bottom panel) targets. The yellow, purple, green and red points show the regions that the nuclear recoil distributions, which are derived from the parameter space of the UMSSM model, can be separable  for  2 $t \cdot y$,  6 $t \cdot y$, 20 $t \cdot y$ and 200 $t \cdot y$ exposures,  respectively, while the blue points refer to that case when this is not possible with our $ \chi^2 $ analysis. The colour coding is detailed in the text. } 
	\label{fig:rate_chi_over_nTOT_Xe}
\end{figure}
In the remainder of our analysis, we investigate that how the ratio between the total number of nuclear recoil events of the Singlino- and Higgsino-like DM neutralinos  changes in terms of different exposures. In this part,  we only use the model-dependent results shown in Figure 7 of Ref. \cite{Frank:2020pui}. The plots in Figure \ref{fig:rate_chi_over_nTOT_Xe} show the ratio between Singlino- and Higgsino-like DM masses in the range of [0.99,1.01] versus the ratio between Singlino- and Higgsino-like DM total number of events between 5 and 50 keV nuclear recoil energy for Xenon (top left panel), Germanium (top right panel) and Silicon (bottom panel) targets. The yellow, purple, green and red points shows the regions over which the nuclear recoil distributions, which are derived from the parameter space of the UMSSM model, can be separable  for  2 $t \cdot y$,  6 $t \cdot y$, 20 $t \cdot y$ and 200 $t \cdot y$ exposures,  while the blue points refer to the case when this cannot be done via our $ \chi^2 $ analysis. The following list summarises the relation between colours and  exposures in  Figure \ref{fig:rate_chi_over_nTOT_Xe}.

\begin{itemize}
\item {Yellow: The nuclear recoil spectra can be separable with an exposure of 2 $t \cdot y$.}
\item Yellow+Purple: The nuclear recoil spectra can be separable with an exposure of \\6 $t \cdot y$.
\item Yellow+Purple+Green: The nuclear recoil spectra can be separable with an exposure of 20 $t \cdot y$.
\item Yellow+Purple+Green+Red: The nuclear recoil spectra can be separable with an exposure of 200 $t \cdot y$.
\item Blue: The nuclear recoil spectra cannot be separable.

\end{itemize}

Clearly, when the total (SD plus SI) event rate is very similar for Singlino- and Higgsino-like DM, separation of these two DM candidates is not really possible, irrespectively of their mass ratio. Conversely, even when the  latter is close to one, separation is indeed  possible even for small event rate differences so long that sufficient exposure is afforded by the experiment. Here, a detector exploiting Xenon would overall be better placed than one using Germaniun or Silicon, as less exposure is needed to achieve a similar level of separation of the DM nature.

\section{Conclusions}
\label{sec:conclusion}

In this work, we have shown that a $\chi^2$ analysis usually adopted in separating DM signals from backgrounds in case of direct detection experiments, when a signal has indeed been established, can also be used to distinguish the nature of the DM candidate.
Specifically, we have shown that Singlino- and Higgsino-like signals emerging from a UMSSM model of E$_6$ origin can be distinguished from each other. While we have shown this to be the case in this model-dependent example, we have also used a model-independent setup to provide a backdrop illustrating the origin of such a difference, using a variety of materials used in such DM experiments.  

Specifically, we have found that varying the SI cross sections largely impacts the difference between the signals in Xenon detectors while it is insufficient in the case of Germanium and Silicon ones. In the latter detectors, instead, varying SD cross sections have a powerful effect in order to discriminate the evidenced signals from the two DM candidates. On the one hand, this means that it cannot be possible to extract the DM nature in Xenon detectors in case of similar scalar and vector interactions. On the other hand, in Germanium and Silicon detectors, the same is true for the DM candidates with similar axial-vector interactions. Hence, different detector materials are differently suited in direct detection experiments in extracting the  nature of a detected DM signal.  

\section*{Acknowledgements}
SM is supported  in part through the NExT Institute and the STFC consolidated
Grant No. ST/L000296/1.  The work of YH is supported by The Scientific and Technological Research Council of Turkey (TUBITAK) in the framework of  2219-International Postdoctoral Research Fellowship Program. The authors also acknowledge the use of the IRIDIS High Performance Computing Facility, and associated support services at the University of Southampton, in the completion of this work.

\end{document}